\documentclass[fleqn,usenatbib]{mnras}

\usepackage[T1]{fontenc}
\usepackage{ae,aecompl}

\usepackage{graphicx}	% Including figure files
\usepackage{amsmath}	% Advanced maths commands
\usepackage{amssymb}	% Extra maths symbols

\title[Women in Science]{Women in Science: Surpassing Subtle and Overt Biases through Intervention Programs}
\author[Bondarescu et~al.]{
Ruxandra Bondarescu,$^{1}$\thanks{E-mail: mn@ras.org.uk (KTS)}
Jayashree Balakrishna,$^{2}$ 
Christine Corbett Moran,$^{3}$ \newauthor 
and Anuja DeSilva$^{4}$
\\
$^{1}$Institute of Cosmology and Gravitation, University of Portsmouth, Portsmouth, PO1 3FX, UK\\
$^{2}$Harris Stowe State University, St. Louis, MO 63103, USA \\
$^{3}$NASA Jet Propulsion Laboratory, Caltech, Pasadena, CA 91011, USA \\
$^{4}$IBM Semiconductor Technology Research, 257, Fuller Road, Albany, NY 12203, USA
}
\pubyear{2018}

\begin{document}
\label{firstpage}
\pagerange{\pageref{firstpage}--\pageref{lastpage}}
\maketitle

\begin{abstract}
This study discusses factors that keep women from entering science and technology, which include social stereotypes that they struggle against, lack of maternity leave and other basic human rights, and the climate that makes them leave research positions for administrative ones. We then describe intervention processes that have been successful in bringing the ratio of women close to parity, compare different minorities in the US, and also consider data from India, Western and Eastern Europe. We find that programs that connect the different levels of education are needed in addition to hiring more women, providing them with basic human rights from when they begin their PhD onwards and promoting support networks for existing employees.
 The authors of this paper hail from Sri Lanka, Romania, India, and the United States. We hold undergraduate and graduate degrees in physics or chemistry from the United States, India and Switzerland. Our conclusions are based on data that is publicly available, on data we have gathered, and on anecdotal evidence from our own experience.
\end{abstract}


%%%%%%%%%%%%%%%%%%%%%%%%%%%%%%%%%%%%%%%%%%%%%%%%%%

%%%%%%%%%%%%%%%%% BODY OF PAPER %%%%%%%%%%%%%%%%%%

\section{Introduction}
\section{Introduction}
Studies show that a child's early interests are not gender specific with females benefiting more from a gender neutral upbringing \citep{raag1999influences,shutts2017early}. The gender gap in Science, Technology, Engineering and Mathematics (STEM) with a significant male to female ratio is then attributable to societal cues as to what constitutes being male and what constitutes being female. Early on these cues are provided in families by the gender specific roles of the parents, and by the toys and activities boys and girls are encouraged towards. The shaming of girls when they act like `boys'€, media depiction of males and females, indulgence of boys when they are exploratory and even break things, while letting girls know that such behaviour is unacceptable for them leads to building societies where STEM exploration and creativity are considered the man'€™s purview. Any challenge to the status quo results in a reaction that is detrimental to women. Studies find that by the age of 6 girls dissociate their gender from cleverness \citep{bian2017gender,bian2018messages}. By the age of 10-11, STEM loses the interest of girls who already believe that a science career is not for them \citep{archer2012balancing}. By highschool there is a significant gender gap. While girls outnumber boys when it comes to taking Advanced Placement (AP) tests, they are underrepresented in advanced placement in STEM fields other than biology and environmental science. Boys outnumbered girls by more than 4:1 among AP tests in computer science, and by more than 2.5:1 in physics AP and 1.5:1 in calculus \citep{ericson2013detailed}.

%because the society is not interested girls associate science with cleverness and see  The STEM gender gap widens 

%, insignificant at elementary school level starts raising its ugly head in middle school and widens in high school. The teen years are when STEM loses the interest of girls because they start to understand that boys and the larger society may not be interested in the STEM girls. % to fit in with their peers start to understand the price they must pay for acceptance. %The teen years are when STEM loses the interest of girls because they start to understand that boys and the larger society may not be interested in the STEM girls. 

%Studies show that young children show similar interests in science regardless of gender. Later, a combination of factors that include the attitude of adults, gender-specific toys and clothes that assume boys are the heroes, the lack of women mentors and/or in leadership roles, and the portrayal of women in the media cause a gender gap. This gap widens as the children grow. The gender gap becomes visible in highschool when science classes are already dominated by boys. 

The gender gap widens in college and graduate school when it becomes common for the women who succeed to stay in science to end up the only ones in their class or research group.  The overall fraction of women with STEM degrees flattens at around 35\%.  The gender ratio varies between the subfields with physics, computer science, and engineering retaining a substantially lower number of women than other fields.  The percentage of women who obtain bachelor degrees in physics hovers around 20\% since 2005 \citep{apsData}. 

A problem that manifests world-wide is that women drop out of science and engineering mid-career. Out of 41\% women scientists, engineers and technologists on the lower rugs of the corporate world, 52\% of them drop out to switch to less technical fields in their mid to late thirties \citep{hewlett2008athena}. In the US in 2017, women held only 24\% of STEM jobs even though there were paid 35\% more than women in non-STEM jobs and had a smaller gender wage gap than in non-STEM jobs \citep{womenSTEM}. This leaky pipeline manifests in scientific research: when one averages over 137 countries, the number of women researchers is only 28\% \citep{huyer2015gender}. Gender parity occurs at masters and bachelor level, but the pipeline start to turn leaky at the PhD level. The current environment weeds out talented individuals of both genders, who leave their fields wondering what they could have accomplished had the environment been supportive.   

%The ones who do obtain a degree in science and try to stay in academia face other hurdles that include a sequence of temporary contracts, where the temporary employee status ensures most rights end with each contract. These 'temporary contracts' can add up to 10 years or more. 

\cite{stoet2018gender} find that countries with greater gender equality have fewer women by percentage in STEM than many countries with higher gender inequality. We present some statistics here and later in the paper explore the subtle biases that cause this discrepancy. It is clear that apparent gender equality in every day life does not mean a gender equal STEM culture. Biases add up resulting in lower creativity and productivity of women in STEM fields compared to  men and in more women than men leaving their field of expertise \citep{2018Report}.  Finland has reached gender parity in natural sciences, mathematics and statistics degrees, while retaining a low number of women in Information and Technology (IT) and engineering \citep{FinlandStats}.  Women hold only 16.3\% of engineering degrees, and 20\% of information and communication technology degrees. The number of women professors in Finland is 24\% across all disciplines even though for the last 30 years, 50\% to 60\% of university graduates have been women. The numbers also vary across disciplines with women in physics comprising 25\% of graduates and only 7\% of professors \citep{banzuzi2013women}. In contrast, women in developing countries are better represented. In the 2016-2017 academic year in India, women obtained around 47.6\% of science degrees, 42.4\% of IT and computer degrees, and 28.4\% of engineering and technology degrees \citep{India2017}. Of the tenure-track faculty in physics 20\% are women, and out of these only 12 women are Fellows of the Indian Academy of Sciences, while there are 197 men \citep{WiredIndiaFaculty}. In the European Union, women comprise 17 \% of the information, communication and technology (ICT) students. Among the member states, Eastern Europe has the highest number of women studying ICT with the highest number of women being in Bulgaria (34.4\%) and Romania (29\%) that is comparable to India's percentage \citep{EuStats}. In natural sciences and engineering, the representation of women in India was 11\% among full professors \citep{SheNumbers}. Thus even in countries with a high number of science graduates, most do not achieve leadership in their fields of expertize.
 
 Intervention programs that focus on attracting talented highschool students to STEM programs in college coupled with a modernization of the course-work were shown to lead to gender parity at college level. The Massachusetts Institute of Technology has successfully brought women's enrollment in its mechanical engineering program to 49.5\% \citep{MITpress}. They find that existent women faculty and students attract more women students. It is reasonable to conjecture that with appropriate intervention programs, the hiring of more women can lead to a snow-ball effect, and that allowing for basic human rights like maternity leave is a necessary part of the process. Programs that connect the different levels of education could perhaps, in time, replace standardized tests. These tests are tools introduced since 1940s that were shown to typically fail at predicting success, while keeping women and minorities out of the best schools \citep{ripin1996fighting,miller2014test}.

Scientists who do succeed in obtaining STEM bachelor degrees and choose to stay in science often work in an array of countries. In the training period that can last 15 years or more their rights are restricted by their immigration status. Obtaining a doctorate degree can take five years or more, and the postdoctoral period extends to the mid (and sometimes the late) thirties. Doctoral students and postdoctoral scholars are temporary employees with very limited rights. Scientists who are immigrants have a restricted visa status and do not have the same rights as regular staff during maternity/paternity leave or illnesses.   More generous rights to maternity leave and childcare are awarded to women who have tenure-track or permanent positions, but such positions are often reached past the child-bearing years. Since the support net is not there when it is needed, most talent is lost at the end of the PhD or after the first postdoctoral position \citep{huyer2015gender}. Some women transition to administrative positions inside the academia, and many quit while believing they were not good enough to succeed, which is an attitude that is not conducive to maximize success elsewhere. 

Unfortunately, the low numbers of women in STEM in the western world are still used as an argument against allocating funds to provide daycare on campus, maternity leave, sick leave and other basic human rights for PhD students and researchers.  Since most women leave their field of expertise in their mid and late thirties, it is likely that the lack of basic human rights does play a role along with the toxic attitude that is naturally exhibited towards minorities.  While women and minorities in the western world choose to stay out of STEM fields whenever they are able to do so because of toxic environment, the inclusion of women should never be presented as a choice between diversity and greatness. It has been shown that collaborations are fully empowered only when there is enough diversity to prevent grouping based on prejudices \citep{page2007making}, which abuse minorities. This ensures social coherence.

To avoid losing our best and most talented individuals mid-career, employers have to encourage a healthy balance between work and family life where paid maternity/paternity leave, egg freezing, sick leave, child and elderly-care are available for all employees independent of nationality, visa status or gender. Today, instead of ensuring all employees have basic human rights, the blame is shifted from the inadequate policies that can and should be changed to the employees, who are accused of not being assertive enough or even worse of daring to "want it all". This culture of blame has devastating consequences for women and minorities, and particularly young mothers who are more vulnerable to depression and self-blame. Instead of finding more reasons for blaming them, we ought to focus on changing rules to help them stay in their field of expertize.

The final goal of society should be to have as many people as possible choosing what they would like to pursue and to be productive and innovative in their chosen fields. No one should be stymied in their endeavors because they have to cater to ordained repressive societal norms. A change in attitudes will only come when a sizeable number of women make the leap. This requires an understanding of the current conditions and support of innovative intervention strategies.

Unlike modeling, sport and acting, science is still presented in abstract, uninteresting ways that have not changed for centuries.  Traditionally, scientists are portrayed as crazy, dysfunctional men who manage to make use of random numbers, strange objects and outdated materials, while the life of actors and models is seen as glamorous and desirable; yet it has its own problems and stories of abuse.   This portrayal together with other choices made for children starting in early childhood discourages girls and most minorities from all countries and cultures from understanding science and engineering. We advocate gender neutral toys, and the exposure of all children to science from the very beginning.  Science can be part of bed-time stories, and part of the curriculum starting in kindergarten and pre-school. Girls and boys should be taught to use their natural curiosity and some of the knowledge accumulated from other people to understand how the world works. It was shown that through intervention programs parity can be achieved even at schools that are highly technical like the Massachusetts Institute of Technology. Furthermore, the degree obtained, whether it is a BA/BS, MA/MS or PhD, should not be seen as an end at which point all interest in the student is lost, but as a new beginning where investment is made towards a smooth transition to the next level.

%ntervention programs do work and that with programs that connect talented highschoolers to college through internships and summer long courses, 
The authors of this paper hail from Sri Lanka, Romania, India, and the United States with formative education in these countries.  All have undergraduate and graduate degrees in physics or chemistry from the United States, Switzerland, and India. This paper discusses gender balance in STEM fields across cultures.  The authors studied, taught and performed research at Cornell University, Massachusetts Institute of Technology, Washington University in St Louis, Harris-Stowe State University, Mount Holyoke College, the University of Illinois at Urbana-Champaign, Pennsylvania State University, the University of Z\"{u}rich, Caltech, the Jet Propulsion Laboratory, the South Pole Telescope, IBM and SpaceX. The conclusions of this manuscript are based on personal experiences, and on data gathered by us/for us and cited references.

%Our children's time is valuable. School should not be something to endure that keeps children busy so that their parents can work.
%This portrayal has significant impact on the younger generation who grow up in front of their tablets, phones and TVs .
%Since the competition is larger, the options in modeling are more limited than if women were to become scientist or engineers.

\section{A first problem: the attitude towards girls and women}
 In most cultures, a woman gets positive reinforcement from an early age only when being passive and nurturing. Her curiosity, analytic mind and ability to take things apart and put them back together are not rewarded the same way. If one goes to a toys store, there are clear indicators that certain toys are for boys, and certain other articles are for girls. From labels to colours one is informed what is for boys and what is for girls. Construction and mechano-spatial toys, trucks, cars all scream `boy'. T-shirts that encourage leadership roles are generally in the boys sector. Only boys can aspire to be batman, superman or spiderman. Girls are mostly expected to dress in frilly and revealing clothes and aspire to be saved by, or if they are extremely lucky, marry the superheroes \citep{graff2012too}. This stereotyping is encouraged by movies \citep{bleakley2012trends}, and books \citep{hamilton2006gender}, and by the larger society as well. It makes children feel unfit if they are attracted to articles from another category. Ideally, toys should be without a gender assigned to them so that parents can let children decide where their interests lie. The stereotyping goes beyond toys. Boys often deconstruct gadgets, reconstruct them, sometimes incorrectly and even break them. This behaviour actually helps them learn and explore and is vital to the growth of STEM skills. Such behaviour by a girl will often be discouraged. She will be admonished and will learn to stop these explorations as inappropriate. This continues in school, college and beyond.
 
The conjecture that attitude and not aptitude is causing the gender gap has been validated by various studies  \citep{penner2015gender, leslie2015expectations}, but no steps have been taken to change the attitude.  From clothes to toys, science and leadership are still sold as a boy-man-interest. Some gender specific books have appeared that recount the adventures and successes of women, but they are by far not enough to change such fundamental stereotypes. 

Instead of continuing to promote a culture of blame that tells women of all ages that they have not succeed because they are lacking, they are not brave enough and not strong enough, we could make changes.  Stores could label most products without assuming a gender. Children products could come with recommendations based on age, size and interests exhibited by the child to allow them to pick in an unbiased manner. Similarly, starting in pre-school all children could be exposed to a variety of science books and experiments, while allowing them to lead towards more depth through asking questions.

More intervention programs that link schools and graduate schools are needed. Such programs could encourage graduate students in universities around the world to go beyond being research tools, and spend some time explaining science to the next generations. This is currently done in one-day workshops that include Expanding Your Horizons Program at \cite{EYH}, and Astrofest at \cite{AstroFest}. These workshops have proven to be extremely successful, and could be expanded to semester long programs that expose children to science.

In addition, women must be made aware that the lack of female role models in STEM and other fields is not because of incompetence but because of a restrictive history. Men have had a much longer history of exploring their passions and interests, of being able to have a say in how life has to be lived, how science is presented, and in being able to vote than women. There was a time when, in the US, women were barred from certain work post marriage \citep{rindfuss1996women}. Women's suffrage happened as late as 1920. Women authors had to write under male names (e.g., George Eliot) because women were not allowed to publish. Such restrictions made it hard for women who were competent to grow and reach their potential. In general it is not just men who prevent women from succeeding. Women who maintain the status quo are lauded by society and also restrict women who think differently. 

During wars, injured men needed care and many women were conscripted into nursing. This history has led the way to women being better represented in biology and health care fields than in other STEM areas. Women must be made aware of their own role in encouraging the success of other women, not underestimating women, and in exploring different ways to promote knowledge for different learning styles. Additionally men must be made aware of the advantages of history that have contributed to their success so that they also become enablers of success independent of sex, color or nationality.

\begin{figure}%[ht]
  \centering
  \includegraphics[width=8cm]{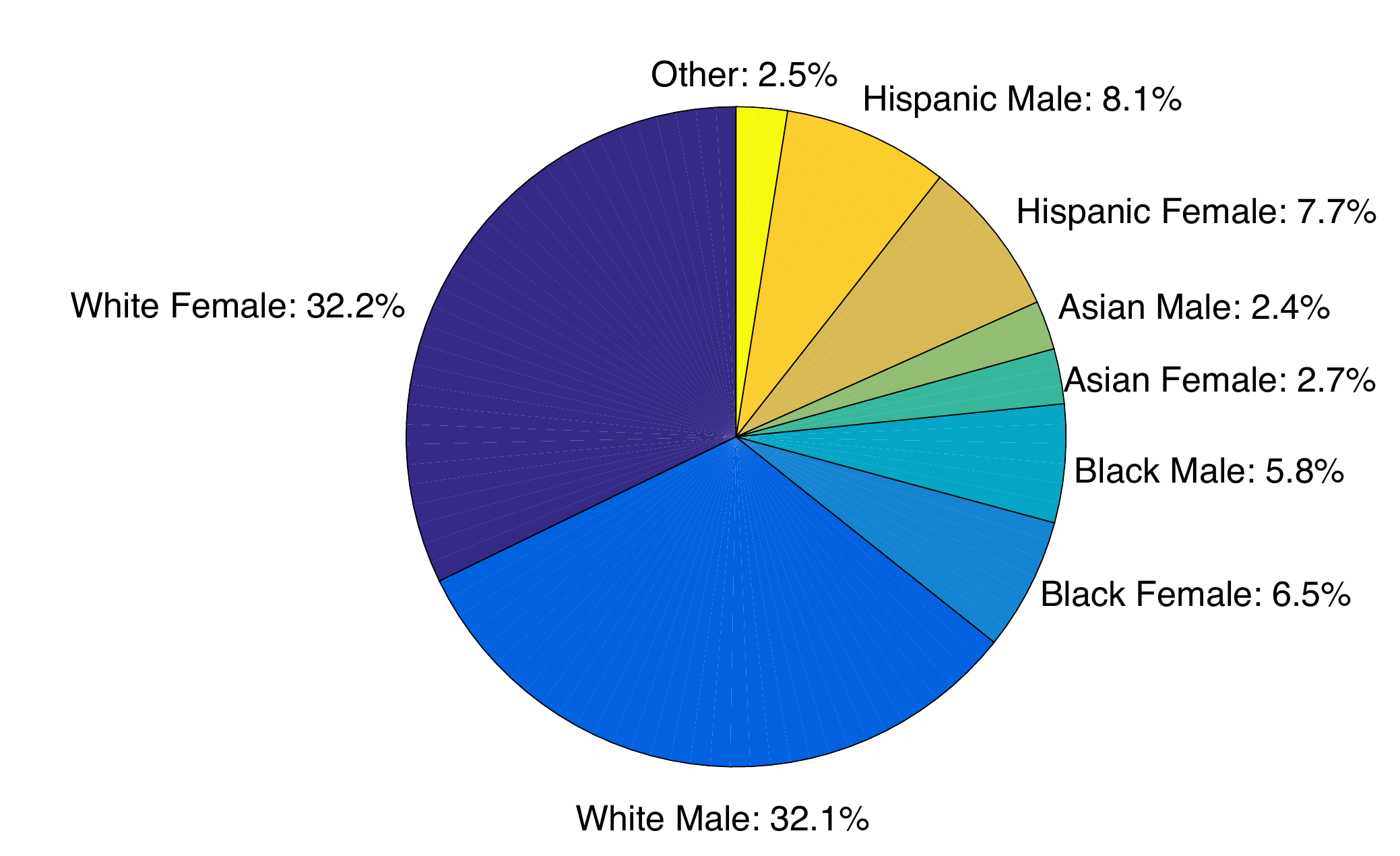}
  \includegraphics[width=8cm]{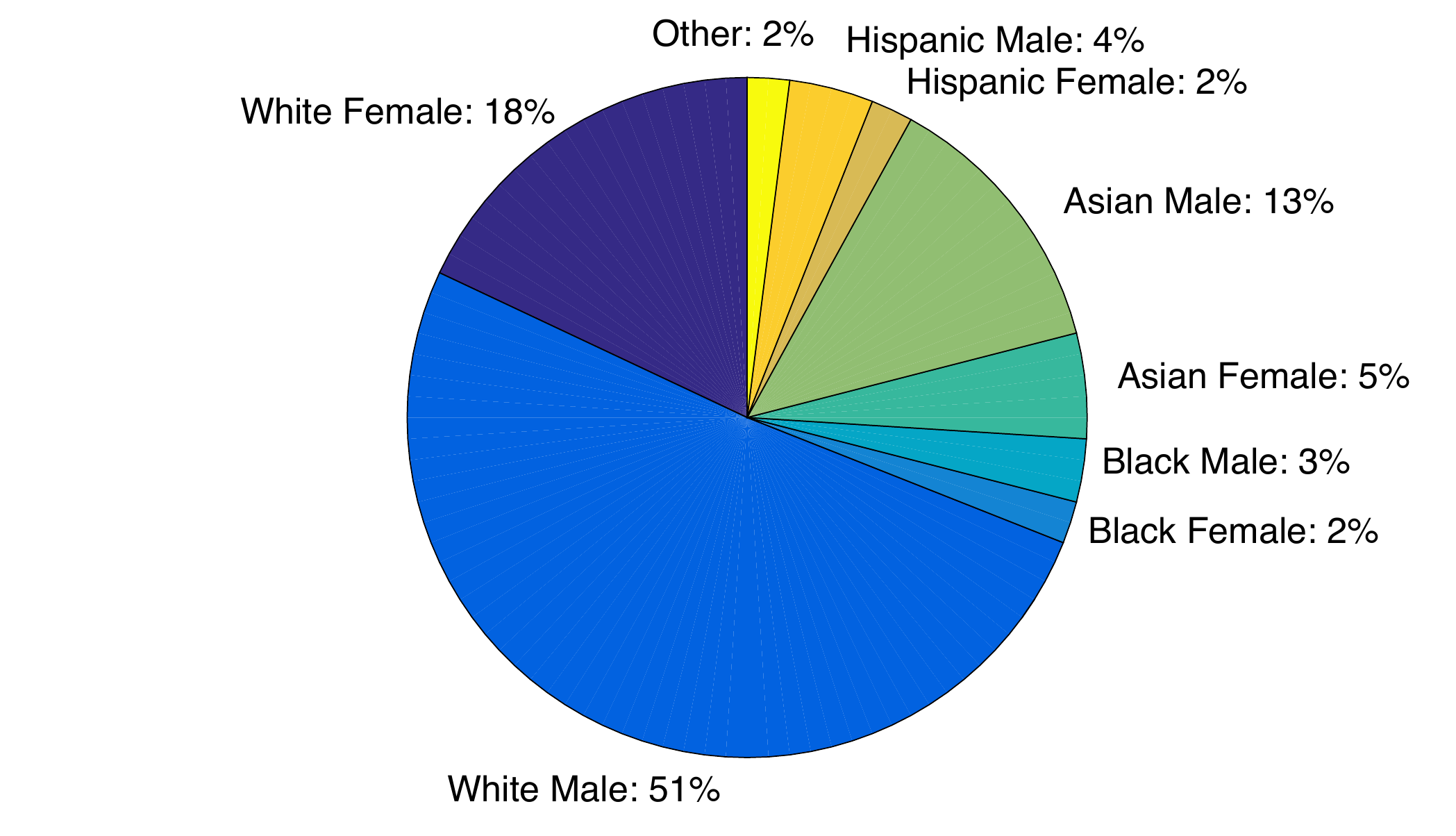}
  \caption{a) US population in 18-64 age group and b) STEM population in the same age group. It can be seen that the percentage of women is always comparable to that of men in the general population, but the STEM population is overwhelmingly male even for Asian minorities where both gender are pressured towards science and engineering. The data is from Guterl, F, 2014.}
\label{fig:pie}
\end{figure}

Figure \ref{fig:pie}a) shows the distribution of the US population of ages 18-64 in terms of race and gender, while Figure \ref{fig:pie}b) shows the distribution within science and engineering. The data is taken from \cite{guterl2014diversity}. In this segment of the US population, white males are 32\% of the population, while making up 51\% of the STEM workforce. Black females are 6.5\% of the population and only 2\% of the STEM workforce. White females are 32\% of population, and also under-represented at 18\% of the STEM workforce. The same trend is observed for Hispanic women.  On the other hand, Asian females comprise 2.7\% of the working population and 5\% of the STEM workforce, and Asian males are 2.4\% of the population, which is a similar fraction of the population as Asian females. They, however, make up 13\% of the STEM workforce \citep{guterl2014diversity}. So, while in the US people of Asian origin are overall more likely to enter the STEM workforce than white people, the push towards STEM is more effective in the male population than in the female population. This data shows that Asian men are almost three times more likely than Asian women to join the STEM workforce. However, the push towards STEM irrespective of the toxicity of the environment, while raising the average STEM competence, suppresses higher end creativity and innovation \citep{nager2016demographics}. 

In developing countries, STEM education is seen as a sure path to a reasonable income. Families then pressure both men and women to succeed in these fields irrespective of other environmental factors. Under pressure, those factors can become something one has to deal with.  The primary and secondary education is focused to this end. This idea is reinforced through school uniforms and standard classroom curricula. Although the society still has a clear bias that favors men, more women make it into STEM fields. A recent study \citep{escueta2013women} found that while women in India experience no bias in school, they experience bias in the larger society. They also hypothesize that there may be biases that are ignored by women, but that this is unlikely. The lack of overt biases may reflect a change in attitude from when one of the authors (J.B.) attended school in India when engineering and technology were considered to be 'smart fields' with boys being suited for these fields. These attitudes are highlighted in the section where the authors discuss their personal experiences later in the paper. Nevertheless, the pressure on boys to learn lucrative skills like computer science and engineering independent of their inclination is higher than on women because they are seen as the supporters of the family. They react to this pressure through increased competitiveness and attempts to eliminate any competition, which makes it more difficult for girls and other minorities to fit in. The pressure to conform to femininity standards and be perceived as non-aggressive shapes women's behavior and makes them leave STEM positions and avoid leadership roles.

%Cultures with group dynamics react differently to the same pressures than cultures which are more individualistic.

For every 200 students getting degrees in India and the US, 52 Indians and 26 US residents obtain STEM degrees. Of these students, in India, 32 will be male and about 18 female (male:female ratio of 2:1), while in the US, 21 will be male and 5 female (4:1 ratio) \citep{STEMDegreesByCountry}. We conjecture that the pressure to be in STEM in developing countries is from families understanding the job market and having a major role in their children's decisions and is not correlated to how comfortable women feel in the field. In the U.S. the career decision comes from individuals who seek out their comfort zones based on the cues they get and from how their aspirations project on the social environment around them.  In attitude studies women in both countries women expressed lower confidence, less assertiveness, and the unlikelihood of expressing an answer for fear of being wrong \citep{Yardi}. 

A lot of learning happens through making mistakes and then reworking through and correcting them. People and societies have been more indulgent to males being exploratory and breaking objects before fixing them.  Women are thus more diffident in exploring because they feel they will be chastised if they do not obtain the right answer on the first try. This attitude results in a significant gap in confidence between women and men. If women are rejected by an employer, they are 1.5 times less likely than men to apply to another job opening by the same employer \citep{brands2017leaning}. Women are generally better at group work because of their attitudes of consideration \citep{Yardi}. However, if the group dynamic is such that the women and men take only the men in the group seriously, it has serious consequences for the creative expressions of women.

The expectation that men do more valuable work than women, and hence are more valuable for the society is far-reaching and goes beyond STEM statistics. In India and China, the sex ratio is skewed towards men, and is still increasing \citep{YouthIndia,hesketh2011consequences}. Parents want fewer children while ensuring there is a son. 

In Eastern Europe, the ratio of males to females is closer to parity in STEM fields. The Information and Technology workforce is about 27\% female in Bulgaria and Romania \citep{Eurostat}.  However, the attitude that women should place men first and be able to do all the house-work still prevails. Females are often expected to be emphatic at work while listening to jokes about beautiful secretaries, shop for office supplies, clean and make coffee at the office, and be the party that buys the necessary things at home from their income alone, while the man of the house enjoys his at a bar. Yet the subtle message that women and minorities are not suited for STEM appears to be more effective in cultures where women are free to choose a non-toxic career in which the employees can be comfortable. 

Cultures with group dynamics react differently to the same pressures than cultures which are more individualistic. In the former, the family pressures to succeed in these fields irrespective of other environmental factors makes those factors something to deal with. However, the existence of such factors in the first place, creates an intimidating atmosphere that can kill creativity and innovation.  In 2017 90.8\% of U.S. patents were granted to men versus 9.2\% to women. The numbers of patents granted to women is still very low even though a positive grow trend of about 18\% is reported: the percentage went up from 7.8\% to 9.2\% in the past 10 years \citep{womenPatents}.

\section{Increasing performance and decreasing abuse through superdiversity}
Minorities stand out, which makes them more prone to abuse by colleagues and/or supervisors. Since they are different, their interests and abilities, and even the way they dress and the way they behave will be noticed and judged. This scrutiny adds additional pressure to fields where very few succeed and pushes out the few who almost make it to the very top of their professions. Any potential abuse is further facilitated by immigration status and by the very limited set of rights provided by universities for their temporary employees, i.e., their graduate students and postdoctoral scholars.

A gender equal society encourages innovation. French (11.7\%) and Russian (15.7\%) female inventors are a long way ahead of Japanese (3.7\%), Korean (4.4\%) and German (5.5\%) female inventors. British (7.3\%) and American (8.7\%) female inventors are relatively close to the worldwide average of 7.2\% \citep{nager2016demographics}.

Immigrants comprise a large and vital component of U.S. innovation: 35.5\% of U.S. innovators were born outside the United States. Another 10\% of innovators have at least one parent born abroad. Over 17\% of innovators are not U.S. citizens, and are nonetheless making invaluable contributions to U.S. innovation \citep{nager2016demographics}. Immigrants born in Europe or Asia are over five times more likely to have created an innovation in America than the average native-born U.S. citizen. Immigrant innovators are also better educated on average than native-born innovators, with over two-thirds holding doctorates in STEM subjects \citep{nager2016demographics}. In part, this may be because there is often a selection process for foreign-born innovators where the ones with the most talent (and perhaps most motivation) choose to come to America because of the significant opportunities this country promises for innovators. 

Women represent only 12\% of U.S. innovators. This constitutes a smaller percentage than the female share of undergraduate degree recipients in STEM fields, STEM Ph.D. students, and working scientists and engineers. The average male born in the United States is nine times more likely to contribute to an innovation than the average female. The United States is therefore missing an enormous potential source of innovation by not creating a gender equal society. Even at this low level, however, the United States outperforms Europe. U.S.-born minorities (including Asians, African Americans, Hispanics, Native Americans, and other ethnicities) make up just 8\% of U.S.-born innovators. However, these groups total 32\% of the total U.S.-born population. Despite comprising 13 \% of the native-born population of the United States, African Americans comprise just half a percent of U.S.-born innovators \citep{nager2016demographics}. Here, too, is an untapped resource of great promise. 

Since many women in science are immigrants with limited visa statuses and human rights, they can more easily become trapped in abusive relationships with their co-workers and academic advisors (or principal investigators). Most subfields have small communities where it can be very difficult to impossible to escape abusers without quiting the field. If they cannot graduate and find a job, which is already challenging with the full support of one's supervisor, an immigrant has to leave the country in shame. This adds pressure and discourages students from seeking help. When they do seek help, they find that complaints (e.g., Title IX) are worded in ways that ensure anonymity for the institution and the abuser while providing little help or protection for the persons being abused.

 Studies find progress is optimal when a super-diversity is maintained, where the teams are not dominated by a single gender or by one or two national identities \citep{page2007making}. They suggest we limit our achievements by limiting diversity. Europe has a strong culture of promoting its own nationals that is partly justified through language barriers. On the other hand, the US, as a new country, included people with a wide range of national identities and  upbringings. Until recently, its openness attracted a diverse scientific community. As a consequence, it still holds the most productive scientific community that exists today. However, the low number of women and minorities, emphasize that the community is not as diverse as it could be, and thus does not maximize creativity. The need for superdiversity has yet to be embraced by scientific communities across the world. %that included the best and the brightest scientists from around the world.

In order to build a gender equal society that taps into the potential offered not just by white men, but also by women, people of color, and other minorities, people should be celebrated and appreciated for their difference in thinking. Typical traits that are today recognized as being feminine such as empathy and compassion need to be recognized as core values in the corporate culture, and not presented as weaknesses. The ability to nurture a culture of compassion, connect on a personal level and coach people needs to be rewarded to enable and retain not only women, but the best and most talented professionals. Work-place approaches that have proved effective in reducing gender bias include (1) gender bias training for all employees, (2) regular round tables with senior women engineers who provide role models for younger women, and (3) special assignments where women can continue to learn new skills while working with mentors outside of their sub-group.

%At first it appears that women originating from countries that focus less on gender equality are more likely to join STEM than their male counterparts. However, data shows that while Asian women are much more likely to join the workforce than white women,
%5.756 x 10^6 in STEM in 2010
% US population 309.6 x 10^6 in 2010
% 0.32 x 309.6 x 10^6 = 99 x 10^6 white men
% 2.94 x 10^6 white men in STEM, about 3% of the general population
% It appears that people who originate from countries who focus less on gender equality like Asia and Eastern Europe are more likely to join the STEM workforce.
 
%In particular, the attitude taken by leaders of institutes and universities is that there is no need to struggle to increase diversity when the institution is already attracting talent and money. This is often stated openly. Unfortunately, studies show that the gender gap in STEM is not diminishing.

\section{Admission and retention in graduate school}
Admissions rely on standardized tests to select their students and top universities boast very high scores. Studies have shown that performance in graduate school is correlated with GPA and with the ability to communicate and interact. The standardized tests, introduced since the 1940s as admission requirements, were found to be better predictors of gender and color than professional success \citep{ripin1996fighting,miller2014test}. These tests play a major role in increasing the gender gap by keeping women and minorities out of top schools. Some graduate schools do not require standardized tests, but recommend them. However, test scores are still the primary basis for evaluating students who are chosen for college and graduate school because they allow for fast comparisons of students of different backgrounds. Programs such as the Fisk-Vanderbilt \citep{miller2014test}, which offer a master that acts as a bridge to the PhD, employ a 30-minute interview instead of a Graduate Record Examinations (GRE) cut-off. It proved to be very successful with a high retention rate of women and minorities of over 80\% towards the PhD. 

Since the late 1990s, Cornell University has opted against applying a GRE cut-off when selecting graduate students in physics and astronomy. If the student has outstanding GPA and research credentials, low test scores are ignored.  Cornell even accepts some men and women with GRE subject test percentiles under 50\%.  In the past, these students did well in their coursework, passed qualifying exams, and graduated without problems. However, most Cornellians have high GRE scores. For their incoming class of 2018 in physics, the average percentile for the Physics GRE Subject Test as reported by students to the school is 78\% for the 5 women who matriculated, and 82\% for the 24 men. The students who reported their nationality were either White/Caucasian or Asian American. The number of students who matriculated with GRE subject test scores under 65\% is 4 or about 14\% of the incoming class. The lowest Physics GRE Subject Test score reported by a student was 42\%, and is held by a young men who comes from an ivy league school and has had a GPA of 3.92. The incoming class has average GRE General Test percentile of 90\% and 83\% for the quantitative and analytic section, respectively. Our brief analysis shows that while one can get away with low GRE scores if they have other outstanding credentials, this is not the norm at Cornell, and that the GREs are still an important part of the selection process.

In terms of gender ratio, out of the students offered admissions in 2018 in physics 23\% were women. Fewer women accepted than men, and so the incoming class of 2018 in the physics department is 17\% women (data provided by the Cornell Physics Department). This is comparable to the average number of women obtaining PhDs in physics in the US, which is around 20\% \citep{apsData}. While these numbers are still low, Cornell has a high retention rate relative to other universities. We looked at data starting 1993, and found that between 1993 and 2011, the Cornell physics program graduated 54 women PhDs out of 74. This means that 73\% of women graduated with a PhD from this ivy league school, when the average graduation rate for graduate students was reported by the National Science Foundation (NSF) to be 59\% in 2008. The high attrition rate is due to the supportive environment. Roles models also play a role. The physics department has 8 women faculty (amounting to 15\% of the faculty body). In 2010, only about 15\% of PhD physics departments had 5 or more women faculty members, 47\% of  bachelor's-granting departments and 8\% of PhD physics departments had no women at all \citep{ivie2013women}. In addition, Cornell hosts about 4-5 women speakers per semester in their physics colloquium. They also encourage women graduate students in proposing and inviting women speakers of their choice, and in meeting with the speaker for informal conversations over lunch and/or dinner. 

In the past decade, the physics department at Cornell admitted an average of 4.7 women/year, and in the decade before that 4.8 women/year. Thus, Cornell data is consistent with the global trend for the advancement of women in science in that physics will not reach parity or increase the number of minority students unless admissions procedures are changed dramatically. In the US, the number of women obtaining bachelor degrees in physics hovers around 20\% since 2005. While retaining as many women from this pool as possible is important, it is not sufficient. An option to increase the number of women enrolled, would be to advertise graduate programs and increase the applicant pool from countries where more women obtain STEM degrees, which include Eastern Europe and Asia. In the long term, attracting more children to science is needed. Graduate students love science and could be the best resource for teaching it to the next generation. If each graduate school had a requirement that its graduate students work with school teachers and teach for a semester at a school in the area, more children would be exposed to science and perhaps be molded into future PhDs. Invariably, teaching in schools would be promoted as a viable career choice, and not be looked down upon as some kind of work that only those who fail to become university professors do. In addition, more universities could have programs where talented highschool students take classes and are part of laboratories for a semester or a summer. Participation in such programs could be offered for free to talented women and minorities. Such a program was proven to work by MIT's Mechanical Engineering Department.

The number of women in tenured professor positions is still very low with some departments having a single woman faculty member and some having no women at all. In industry, the percentage of women-headed ventures flattened at 17\% in 2012 \citep{teare20172017}, and has seen no growth since. These low numbers emphasize the need to (1) invest in intervention programs that work (2) hire more women to increase the number of women mentors and potentially induce a snow ball effect that attracts more women, and (3) to build support networks that connect existing students and faculty (a) among themselves and (b) to the next generations. Universities could work closely with schools to expose more students to STEM and  with companies to make a smoother transition for their graduates, while providing support and basic human rights to ensure more women become leaders, mentors and faculty and increasing the number of women faculty and permanent staff. Since only a few percent of PhD holders become professors or permanent staff \citep{larson2014too}, it is important to stop seeing the degree as the ultimate goal that in itself insures success and invest more in connecting STEM graduates with potential employers via workshops on campus and through summer and semester-long programs. This is currently done only in certain experimental fields where the expertise obtained in graduate school is directly relevant for acquiring patents while theorists and other experimentalists fend for themselves. 
% For women faculty members, feelings of isolation, lack of respect of colleagues, and difficulty in integrating family and professional responsibilities are major factors in attrition from university careers. \cite{committee2007beyond}
% PhD granting departments with 5 or more female faculty: https://www.aip.org/statistics/reports/women-among-physics-astronomy-faculty
 
\section{Intervention programs and single Sex Education: part of the solution}

It is clear that intervention programs that address the gender gap do work. In the US in the 1990s boys outnumbered girls in science and math classes in high school, but the numbers have evened out and in some cases girls outnumber the boys \citep{highschoolNumbers}. %https://edsource.org/2017/girls-now-outnumber-boys-in-high-school-stem-but-still-lag-in-college-and-career/578444
According to this study intervention programs like A AUW, the Girl Scouts, Girls Inc., Tech Bridge and Girls Who Code,  that offered after school STEM programs and scholarships for girls were a big part of why this change occurred.  This clearly shows that the gender gap is not due to innate abilities of men versus women in these fields.

The pathway to interest in any field is often through activities one grows up doing. A study of video games mentions that despite an increase in female characters, games still depict them often in secondary roles and sexualized them much more than their male counterparts \citep{lynch2016sexy}. Stereotyping of roles starts with children'€™s toys with clearly defined gender roles inherent in them. These have to be removed to allow children to choose the roles they play.  Shops, games and shows have to portray women and girls as valuable assets of the society, and not as sexual objects. Furthermore, the role of the media, which is led by educated individuals, should not contribute to the objectification of women and should not  pressure women to conform to some fake ideals.

While early learning of girls has to be more supportive to toughen them up to stand up to the rigors of male dominated fields, there is a need to remove at least some of the toxicity of the STEM environment. Women and minority students should be and can be taught to take risks, make mistakes and be bold similar to how boys are typically encouraged. For this, mentoring has to happen from kindergarten age onwards and continue to evolve with age. The love of learning and the growing of skills needs to be a continuous focus especially post-graduate school. Though attaining the PhD may seem as the end goal, it is just the beginning of a professional STEM career.

The percentage of women in physics, engineering and computer science in many universities is still of the order 20\% or under. This means the women attempting to make it will likely be the only female in their research group, and sometimes the only woman in the class, which increases their vulnerability to abuse. Women in science are more likely to be in medicine and biology rather than physics and engineering fields. Male students are more likely than female students to take engineering (3\% versus 1\%) and computer science courses (7\% versus 4\%) and are enrolled in AP computer science at much higher rates (81\% males; 19\% females) \citep{NGC}. %€˜https://ngcproject.org/statistics. 

The larger number of girls in science and math classes has so far failed to lead to an increased female STEM enrollment in college. This emphasizes the need for intervention programs that connect highschool students to colleges, and of more women faculty, graduate and undergraduate students to whom highschool and college students can relate, see them as proof of success and as role models. Such programs were proven to work. 

\begin{figure}%[ht]
  \centering
  \includegraphics[width=8cm]{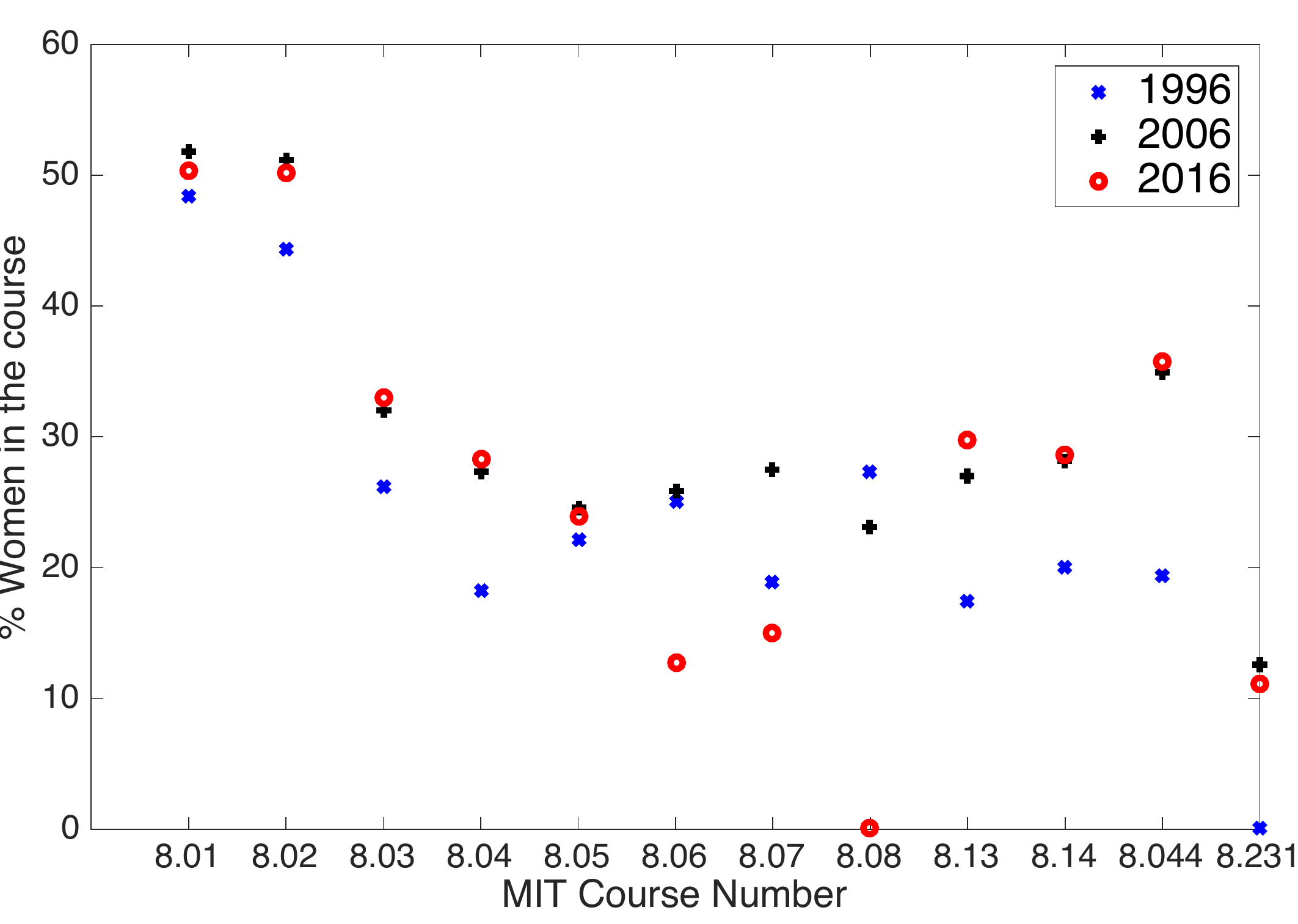}
  \includegraphics[width=8cm]{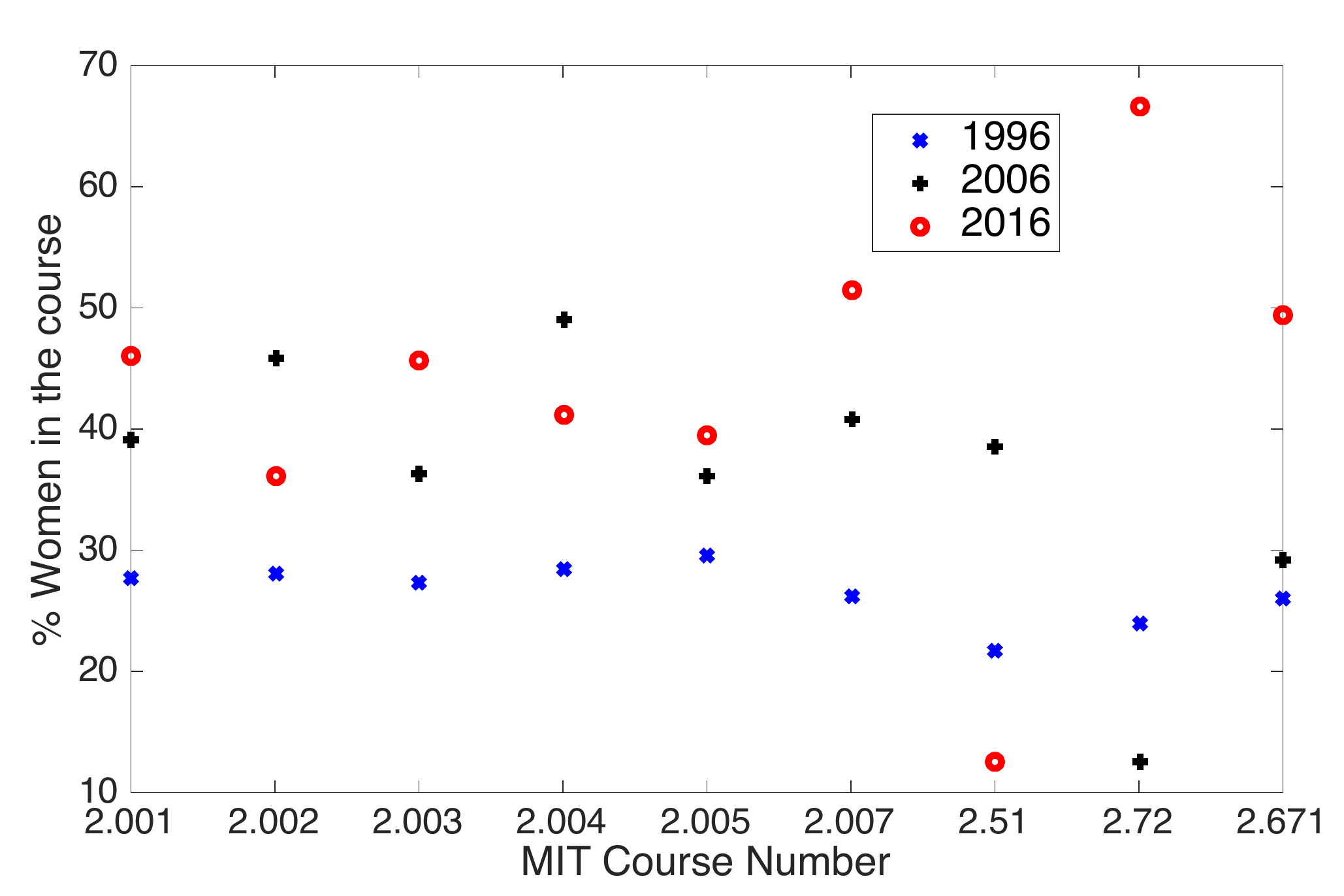}
  \caption{The percentage of women at MIT in a sample of a) Physics b) Mechanical Engineering courses. In physics, parity is only reached in the first two introductory courses, while the rest of the courses held a mean of 20\% women in 1996 and 22\% women in 2016. In mechanical engineering, in 1996 the mean was 26\% women, but in 2016, after a successful intervention program and after a modernization of the curriculum, the same courses have a mean of 43\% women.}
\label{fig:MITWomen}
\end{figure}

In a climate where women receive 19.5\% of bachelor degrees in engineering  and only 7.9\% of mechanical engineers are women, the Massachusetts Institute of Technology (MIT) succeeded to attract 49.5\% women in 2017 to its mechanical engineering program. Fig. \ref{fig:MITWomen} shows the percentage of women in a sample of 12 physics courses and 9 mechanical engineering courses from MIT. The data is taken from the interactive map available online \cite{GenderDiversityMIT}. From the figure, it can be seen that mechanical engineering department went from under 30\% women in 1996 to almost-gender balanced classes.  In physics, in the sample of courses we considered, gender parity is only reached in the first two introductory courses. The rest of the courses held a mean of 20\% women in 1996 and 22\% women in 2016. In mechanical engineering, in 1996, the mean over the courses considered, was 26\% women. The situation has improved dramatically by 2016. After successful intervention programs that increased the number of women admitted and modernization of the curriculum, the same courses have a mean of 43\% women. It is notable that the number of male students that enroll in the courses has increased as well. %Introductory computer science courses show a similar increase in the number of women enrolled.

Gender parity was achieved through deliberate structural changes of the department that included (1) stimulating talented women to apply to MIT through on-campus visits and intervention programs like the Women's Technology Program (WTP), where talented highschool students live on campus and participate in classes and laboratories for an entire summer, (2) increasing diversity through the hiring of more women faculty, and (3) modernizing the curriculum in both content and pedagogy \citep{GenderDiversityMIT, xugetting}. The latter part should not be overlooked. Courses can no longer push students to simply acquire information that can already be found on the Internet. They need to instead focus on teaching students to understand and use available data and tools. The presence of women faculty was shown to play a crucial role in attracting other young women to the department who saw them not just as role models, but as a proof that success and a stable job can be attained in the field as a woman \citep{xugetting}. 

Another tool are student blogs that show that the campus is no longer predominantly male and that women can thrive at MIT \citep{xugetting}. They promote their experience through social media, which attracts more students from the next generations. Students participating in laboratories and classes on campus took the social media by storm when they shared selfies of the experience that lead to the hash tag \#ILookLikeAnEngineer.  They even proposed a reality in which women scientists are revered as much as actors and athletes \citep{GenderDiversityMIT}.

Another resource are single sex high schools or women's colleges for undergraduate studies. They can substantially help women pursue and command STEM fields. In the absence of boys/men, the focus is purely on women's capabilities and not contrasted against the opposite sex. Especially in cultures where boys are regarded as more valuable from a young age, having single sex schools can help to mitigate that effect in women's daily lives. Though girls grow up with the added burden of having to be more pleasing in terms of physical beauty and subservient in attitude, they learn to be certain of their aptitude in academics due to single sex education. It also provides an environment for girls to be mentored and recognized in the absence of boys which can have a lasting effect as they grow into young women.

Mount Holyoke is the first of the seven sister colleges to be established in 1837 when Ivy league colleges were male only and still continues its mission for leading the way for women in STEM. From 1966 to 2004, according to the NSF's Survey of Earned Doctorates, Mount Holyoke graduated more women than any other liberal arts college who went on to get U.S. doctorates in the physical and life sciences (356 and 109, respectively). This puts Mount Holyoke in the top 2 percent of all colleges and universities--even major research universities with at least double the enrollment and faculty. Among all colleges and universities, Mount Holyoke ranks eighth (tied with Stanford and Wellesley) in the number of graduates who earned U.S. doctorates in physics from 1966 to 2004; ninth in chemistry; and sixteenth in biology.  Mount Holyoke also leads with its commitment to minorities. From 2000 to 2004, Mount Holyoke produced more international (non-U.S. citizen) female graduates who went on to receive U.S. doctorates in the physical and life sciences than any other college or university. Twenty-three MHC alumnae received U.S. doctorates in life or physical sciences, compared with 21 women from the University of California-Berkeley, 19 from Harvard, and 17 from the Massachusetts Institute of Technology. Among those 23  Mount Holyoke alumnae who went on to receive U.S. doctorates in life and physical sciences from 2000 to 2004, 22 of them are minority women, the highest number along tier 1 liberal arts colleges in the United States \citep{MHC}. 

%\begin{figure}%[ht]
%  \centering
%  \includegraphics[width=8cm]{HBCUFig.pdf}
%  \caption{The number of black students enrolled in STEM is shown for every 1000 black students in accredited U.S. bachelor's degree programs in the Fall 2008 semester. It can be seen that the gender ratio is close to parity only in HBCUs.}
%\label{HBCUsFig}
%end{figure}

Similar to women's only colleges, the role of Historically Black Colleges (HBCUs) is to reduce the isolation of minorities in Predominantly White Institutions (PWIs) \citep{reid2012women}. In polls on HBCUs versus PWIs, it is clear that black students on an average felt better supported in HBCUs than in PWIs. \cite{seymour2015grads} reports that 29 \% of black graduates who did not attend an HBCU said they were ``thriving in financial well-being,"  and 51 \% of black HBCU graduates reported doing so. Most HBCUs reach gender parity in STEM courses and in some female enrollment exceeds male enrollment \citep{simms2014educational}. The lack of male students is connected with the hardships they endure while living in underserved communities \citep{cuyjet1997african,white2013black}. Black people of both genders have a longer history of exclusion than white women \citep{hine1997hine}.
%Racial minority issues have a different dynamic than gender issues. Underserved communities are underserved right from the beginning while women in STEM seem to have this happen post middle school. Nevertheless there are parallels: role of women's colleges and HBCUs, history of exclusion, need for intervention programs etc. 
%Gender parity is less of of problem there. In HBCUs the ratio of black women: black men is reported to 19.2:23 while the same ratio is about 1:2 in all institutions and 1:3 in for-profit institutions.
%https://www.insidehighered.com/news/2015/10/28/survey-finds-big-differences-between-black-hbcu-graduates-those-who-attended-other

Unfortunately, even within women-only colleges and HBCUs fields that are associated with being ``smart" have less representation than fields associated with ``perseverance". In Mount Holyoke College, the Physics department (5 faculty in physics, and 2 in astronomy) is clearly smaller than biology (13 full time faculty and 2 visiting lecturers). Wellesley  also has far more students in biological sciences than in physics \citep{Wes}. Some HBCUs have comparable biology and physics departments, but most do not, e.g., Morehouse College is a male liberal arts college with 12 biology faculty and 10 physics and engineering faculty, while Xavier University has 12 biology faculty and 5 physics full time faculty. Another issue is that, in many colleges, a significant fraction of the teaching is done by adjunct professors who are underpaid and have fewer rights that full professors. %https://www.wellesley.edu/oir/factbook17/fall-enrollment-by-declared-major

Studies say girls disassociate their gender from smartness from a very early age \citep{bian2018messages}. Minority students who are also targeted by negative stereotypes do the same. This is disturbing on multiple scales since not only does it keep talent from certain fields, it also makes people not trust themselves and stops them from reaching their potential. They end up not believing in their own strength and creativity, which leads to less innovation and a lack of belief in the smartness of other women/minorities. 

\section{Mothers in Science challenge the maternal wall}
The subtle pressure in STEM is to dedicate one's life to work, and to feel guilty when taking time away from what's considered to be one's life-long passion. This leads to postponing having a family sometimes indefinitely. When such pressures are defied, it can be assumed that the person will quit the field sooner or later unless they are already established and that they cease to be a worth-while time investment. The same can happen when an illness appears. So, if the decision to have a child is made, the tendency is to hide the pregnancy for as long as possible, and to avoid mentioning it until after a job offer is made when applying for jobs. The child will then be planned between jobs or assignments if at all possible so that the pregnancy does not show on one's CV. Beyond this, the prevalent attitude is that funding maternity leave and providing adequate day-care will not increase the number of women in the STEM workforce. It is accepted that maternity leave and day-care on campus should, in theory, exist and be available, but also that there is no need to hurry and make any changes when such services are not available because changes that lead to providing mothers in academia or industry with adequate support will not matter. This attitude, while far too common, it is, of course, faulty and endangers lives.

While exchanging work duties with colleagues and planning children in vacations may be possible, it should not be the norm.  Studies find paid maternity leave is not a luxury. It is a basic human right that can be life saving for both mother and child. The lack of antenatal leave has been associated with a three-fold increase in the risk of pre-term delivery and has been observed the have similar effects on the birth-weight as smoking during pregnancy \citep{ceron1996risk,del2012intrafamily}. Mothers who take short leaves are up to four times less likely to initiate breastfeeding when compared to those who do not work \citep{huang2015paid,baker2008maternal}. Paid leave for mothers is linked to increased breast-feeding rate and returning to work is cited as the top reason for breast-feeding cessation \citep{schwartz2002factors}. The American Academy of Pediatrics says infants not breastfed face more than 3.5 times the odds of Sudden Infant Death Syndrome (SIDS) mortality when compared to exclusively breastfed babies \citep{eidelman2012breastfeeding}. Furthermore, paid leave has long-term benefits that go beyond survival of the child, and benefit the child's long-term development and leads to higher achievements \citep{berger2005maternity,carneiro2015flying}. It is noteworthy that paternity leave is linked to an array of long-term benefits as well \citep{huerta2013fathers}.

Many scientists are immigrants for at least some part of their career and thus not eligible for standard maternity leave and child-care support offered to citizens, which puts them in a particularly vulnerable situation even in countries where such leaves are the norm. They have to rely then on company/university-specific policies that also provide full support only for permanent employees. Graduate students and postdoctoral scholars are temporary employees with limited rights that end with their contracts. Yet graduate school lasts an average six years in the US and four years in Europe. The postdoctoral period can still last ten years. This system leaves men and women in their twenties and thirties unprotected for maternity, paternity or illness. In companies, when policies for paternity and maternity leave exist, their applicability depends on the project the person works on with successful employees that are difficult or impossible to replace being constraint to have less time off.

In the academia, most universities offer no adequate support system for mothers to be until they become professors, which often happens in their forties. At that time it can be too late to start a family.  Even for permanent staff, maternity leave can be complicated to attain e.g., at Penn State University maternity leave is comprised of gathering sick day leaves.  Furthermore, waiting to provide support until the position is tenured or tenure-track gives the message that one has to postpone having a family, and often choose between attempting to have a career that is likely to not work out and having a family, which adds to the toxicity of the environment.  For students and postdoctoral scholars, the work schedule is flexible and an understanding advisor will try to navigate the system to help their postdoctoral scholars and graduate students take advantage of various workarounds. Unfortunately, the attitude towards maternity coupled to the lack of supportive services for child rearing, forces women to either postpone childbearing until their career stabilizes, which could be never, or jeopardize their career by hitting the maternal wall. Since many scientists tend to be immigrants, they will not have the right to maternity leave or sick leave even in countries where such leaves are the norm. 

The few specific programs that encourage women to return to work are done in a proforma way. For example, the Horizon 2020 program of the European Union included the Marie Curie Career Restart (CAR) Program to decrease the gender gap and help scientists who were out of the field come back to work. In physics, it had a lower success rate than the regular program; so experienced staff advised applicants against checking the CAR box when they qualified. Scientists were told this was because all the money came from the same pocket since the CAR program had no additional funds.  Switzerland funds programs that aim to fix the leaking pipeline to encourage the hiring of more women faculty. However, departments and advisors neither have the requirement nor are encouraged to report pregnancies to human resources or provide maternity leave beyond the end of the contract. When asked Human Resources argue that they try to provide a case-by-case workaround when the pregnancy is reported early enough to their office. Advisors are generally supportive. This means they tacitly accept a lesser presence in the office during the contract while pregnant, fewer research results, and perhaps a longer time to graduate. Still, at the end of the contract, women scientists can end up in their home country unemployed, heavily pregnant or with a young child, and without the right to paid leave or even health insurance. Unfortunately, in these circumstances, the chances of staying in science are minimal. The women who are under-contract and have children, have to plan and get on the waiting list for the often extremely small day care on campus sometimes before their pregnancy even begins. The insufficient child-care facilities are a problem that is common to university campuses across the world.

Many women who have shaped our world have had and raised children.  Marie Curie raised two daughters -- her eldest won the Nobel Prize; the other was a bestseller author and journalist with a leading role in UNICEF. Marie Curie, her daughters, and sister were leaders in a male-dominated society and excelled. They succeeded in spite of the society they lived in. Marie chose to not only work hard, but to share her life with a man of outstanding moral uprightness. Pierre Curie was already established in his field, and did not conform to recommendations. He went as far as refusing the Nobel prize until the award included Marie. Marie Curie relied on the support of her father in-law to raise her daughter while she worked in the laboratory. Short maternity leaves and lack of available childcare on campus, force women scientists to rely on similar solutions today. Yet the extended family support system is shakier than in the past. Retirement has been pushed back, and if a grandparent is healthy, they will often work themselves instead of carrying for grandchildren. People also choose to have children later when their parents may be too old to help and in geographic locations that are far from where their family is. 

Albert Einstein and Mileva Maric were graduate students when they started their family. They were also both immigrants. Z\"{u}rich's Politechnic institute did not support them. Einstein received his PhD, but could not find work within the university. After searching for work for a long time, he was hired by a patent office. While Albert searched for work, Mileva returned to Serbia where their daughter Lisserl was born and is believed to have died or been given away a year or two later after she became ill. Mileva never received her PhD. Today, infant mortality is still correlated with lack of maternity leave and stress during pregnancy. Later Mileva worked with Albert to produce relativity while receiving no credit for it, and was quoted for saying that they were but one stone, and that with fame often one takes the pearl and the other the shell \citep{Mileva}. 

Below we quote a series of prominent scientists regarding motherhood in science. The Head of the Gender Equality office at the Swiss national science foundation writes "I am myself a mother of two kids and left research for this reason."  A colleague with four children who is a prominent scientist at an ivy league university in the US comments "I have never had a paid maternity leave, except for my first child in which case it was not official. My adviser was extremely supportive and I keep thinking of him with the highest respect. For the other three kids I was in transitioning between cities/positions and planned so that the break would not show on my CV. Financially it was definitely negative." She succeeded to stay in science because her parents and in-laws took turns in providing child-care. Another colleague was cautiously optimistic "Things are getting better slowly in terms of maternity leave both in Europe and the US, but this does not mean perceptions in physics with our male colleagues are changing as fast. I waited to have a baby after I was 35 and I wouldn't recommend this choice (coupled with the risk of health problems and fighting against your biological clock). The main reason was that my husband and I were doing long distance [commutes]/not in stable positions". A Cornell PhD who has returned to India with her husband when he obtained a professorship there writes " I have a 3 year research grant from the government, but after that things are again a bit up in the air. The money for this 3 year cycle has also not been sanctioned - so all my work currently is unpaid ! Pay or not, academia is pretty ruthless so, one has to keep working" (she had a 7-months old daughter at the time). 
% Instead of making parents feel guilty that they dared to have children, and can no longer spend all their energy on work, we ought to change the rules and help them fit back. for adequate human rights to all employees including graduate students and postdoctoral scholars, who spend many years under the temporary employee label with preciously few rights. 

 To remove some of the toxicity of the STEM environment, employees should have access to at least 40 weeks of maternal and/or paternal leave \citep{ruhm2000parental}, and have sick leave, childcare and elderly care available on campus. It should not be the norm that women forego having children to advance in a field or wait until their career stabilizes and it is too late. Combined child-care and elderly-care facilities could exist where the children of employees can participate in various activities with their grandparent or with another elderly person. While such policies might indirectly decrease the gender gap, their primary role would be to enforce basic human rights. 
 
 % will make it easier for employees to work when having a family and to contribute to mentoring the next generation when their health fails,

{\bf Ruxandra Bondarescu describes highschool in Romania, college at the University of Illinois at Urbana-Champaign (UIUC), graduate school at Cornell University, and mixing postdoctoral studies at Penn State University and the University of Z\"{u}rich (UZH) with motherhood.}

I attended the Grigore Mosil Highschool in Timisoara. It was considered the best highschool in the city for studying STEM. All students had the same curricula and had to study informatics for 8 hours every week, mathematics for 5 hours and physics for four hours a week in slight detriment to general education (e.g., music was not taught; geography, history and biology were reduced from two to one hour a week, etc).  The 30 students in my class happened to be equally divided between sexes: we were 15 young women and 15 young men. Out of these students, three went to college and graduate school in the US and all happened to be female. Today, the class graduating in 2018, has 141 students (which are divided in 5 groups or classes) and is about 45\% female. When I went to Z\"{u}rich as a Dr. Tomalla postdoctoral fellow, the only female faculty in Particle and Astrophysics at UZH was a graduate of Grigore Moisil. 

My bachelor degree is from the University of Illinois at Urbana-Champaign. As an international student in a state university, it was difficult to obtain funds to pay for tuition. I was a full time student for only one year when I was awarded a teaching assistantship that included a tuition waiver. The assistantship was obtained with help from physics advisor Linda Lorenz and office-mate Galina Wind. I was also working in research at National Center for Supercomputing Applications (NCSA) with a supportive advisor. At the graduation ceremony, we were 4 women, which constituted about 14\% of the physics students graduating in 2003. Of my class, I was the only woman going onto physics graduate school. Today, among the faculty, the UIUC physics department has 9 women out of 57 professors, and is 15.8\% female. 

I then joined Cornell University's physics department and graduated five years later. I was one of the 6 women entering physics graduate school that year out of 40 students. I started on a two year fellowship with the support of the department head who ended up being one of my two thesis advisors. I support the multiple advisor system. My two advisors made a superb team that helped me stay in science and excel. From the women colleagues in my year: 4 graduated with a PhD, one took personal leave and never returned, and one was a transfer student who obtained a Masters from Cornell and went on to finish her PhD at her home institution. Three of those four women who succeeded to obtain a PhD in physics from Cornell went on to pursue STEM careers, and two are still working in STEM today. 

Cornell's physics and astronomy departments are known across the US as non-toxic and particularly supportive of their students. Over a 20 year period, the graduation rate is 73\% for the physics students who are women, which is higher than national graduation rate of 59\% for PhDs over all fields. As part of the graduate women in physics group, we received funding from the graduate school, and from the physics department to invite one woman speaker of our choice per semester for the main physics colloquium and to meet with existent female faculty and speakers for lunch or dinner. We were given the opportunity to propose and invite leaders in the field for seminars. This was particularly empowering for graduate students. We also had a panel meeting per semester where Cornell's few women faculty would give us advice on how to succeed.  Beyond this, all Cornell graduate students can apply and receive funding from the graduate school to partially fund travel to conferences, which, combined with funds for students from APS, made my presence at all major meetings in my field possible. Today, the physics department at Cornell is 17\% female boasting 8 women professors out of the 45 tenured and tenure-track faculty. The number of women professors is still low enough that in the five years I was there I have never taken a science class taught by a woman and I have taken over 20 courses in graduate school. This was also true for my undergraduate period at UIUC.

After completing my PhD, I started a postdoctoral position at Penn State University. Two years later my first child was born. My postdoctoral advisor was extremely supportive. He worked with staff to find ways to pay me for the next year and provide leave for me. The H-1 B visa requirements were of such a nature that he had to explicitly write how unqualified I was to be able to keep me on because postdoctoral scholars are at the bottom of the salary bracket. I had a month of unpaid leave, and was encouraged to come to work only for meetings and to bring my child to work. My salary was also increased. The day care on campus had a long waiting list, and I ended up enlisting the help of extended family members to care for my son. I also had a private office where I could breast-feed if needed. At Penn State, we were encouraged to invite speakers for the seminar of the center there, go to lunch and dinner with speakers, and funding was available to support one or two yearly visits from collaborators. This helped me write articles with colleagues at Caltech, the University of Mississippi, and Syracuse University. As the only woman in the institute at the time, I had the opportunity to be at dinner with primary donors for the university, and I also led the team of postdoctoral scholars and students in presenting our research to representatives of the National Science Foundation.

I was next awarded a postdoctoral fellowship at the University of Z\"{u}rich (UZH) that lasted five years. When I arrived with my one-year old son, I first tried taking my son to work with me. This was not feasible since the office was shared and it was disturbing to my colleagues.  The small day-care on campus was very friendly, but had a  waiting time of more than one year after registering with priority given to permanent staff (postdoctoral scholars are temporary employees). So, my mother retired from her job as a doctor, and came to help care for my son. The environment was supportive. I was allowed to invite one or two collaborators per year for the Particles and Astrophysics Seminar at UZH, and in my last year, I received partial funding to give invited seminars at universities across the US. My research at UZH was featured several times in various journals across the world, which included the IEEE Spectrum, the R\&D Magazine, New Scientist, and the MIT Technology Review. The professor I worked with was praised by the university for our joint work, and, while I was there, he was promoted to the US equivalent of associate professor after 20 years of being an assistant professor, which is tenured in Switzerland, but comes with lower pay. 

When my UZH position ended, I was eight month pregnant with my second child. I gave birth about 6 weeks after my appointment ended, and as temporary staff I had no right to maternity leave since my appointment would have ended at that time even if I had not been pregnant. When I returned to my home country without health insurance and unemployed, I explained to my mother that I had been temporary staff even though I had been employed for five years in the same place. My mother saw this as a violation of not only my rights, but also of those of my child, which should not happen, and was outraged as a doctor, a woman and a mother by the rules that are commonly applied in STEM.  So, eventually, I gathered enough courage to contact Human Resources (HR) after encouragement from a program for fixing the leaking pipeline who had no funding to help, and I was told that indeed if the delivery is 6 weeks or more after the contract ends, the university has no obligation towards temporary employees. However, they said that if I had I reported the pregnancy early enough they might have been able to find a work around. They also emphasized that departments and advisors have {\it no} requirement to report the pregnancy of their employees to HR. I was then advised to contact the Swiss National Science Foundation. The head of the equality office personally responded to my questions and said there was no funding she could access for such situations, and that she quit science because she has two children herself, but was optimistic about the future. Overall, it was clearly explained to me that it was my fault that nothing could be done for not reporting the pregnancy early enough and for not investigating a potential work-around. I was also told I could have legally staid unpaid in Switzerland for three months after my contract ended, and since I chose to not avail myself of this opportunity, it was my fault for potentially being without health insurance elsewhere. I still stand by my choice since I do not think I would have found a job with a newborn child in those three months and to return home with no savings, a baby and young child would not have been a better solution.%I could have legally stayed unpaid in Switzerland for another three months, continue to pay for health insurance and for the living expenses from my savings, give birth there, and then returned to my home country with a baby and a young child. 

I was the second woman to give birth from my group. The other colleague was in a similar situation where no extension or maternity leave was awarded. Her position had only been one year long. She asked our advisor and the program officers if she could get an extension, but when the answer was `no' she never followed up with HR. She was not able to come to work often because she had a high risk pregnancy, and her previous child died a few days after birth. She had also suffered a number of miscarriages before that. It was assumed that as the only other woman in the group I could advise. When she looked ill, I sent her to the hospital, and emphasized the importance of prenatal care over work. Once she delivered her baby there was no available day-care, but her husband was eventually allowed to come from Pakistan to help. She did finish a technical article based on work performed at UZH after returning to Pakistan. She told us the environment was less toxic in Switzerland than in Pakistan where her female colleague was not allowed to skip work to be with her very young child, and often came to work crying. 

{\bf Prof. Jayashree Balakrishna discusses her education in India, graduate school in the US, and teaching experience at Harris Stowe State University.}
My education started in India. My father was an engineer and my mother stayed home to raise my two siblings and me on a single income. When I graduated high school my field was STEM-biology (Math, Physics, Chemistry, Biology, English).  In my class there were about 24 boys and 16 girls. There was a definite bias from faculty and students that males were better at math and engineering that they had no qualms displaying. People were comfortable with women wanting to be doctors, but not engineers. This bias was also exhibited by some of the women. In the STEM-engineering class (Math, Chemistry, Physics, Engineering-Drawing, English) there was only one woman and the rest were men. She ended up studying English in college. There was an option of choosing domestic-science instead of Mathematics even in the STEM field. Two women took this option from my class and one of them ended up becoming a doctor. Two male student in the next batch took that option and created a hue and cry in the school with even some of the women saying they did not know what they were doing.  

However, at home the expectation was that STEM is important and that one should do STEM. Nobody at home could change prevalent attitudes and so they would not empathize with the situation. You were supposed to play the best hand you could with the cards you were dealt. It always seemed that guys in class could talk to each other and bounce off ideas but the few women aspiring for Math-Engineering glory were isolated.  Since it was a preposterous idea for a woman to think she was able to excel in these fields, this meant the women in STEM tried to avoid the public eye. Some even seemed to underperform for fear of offending the norm. I attended a women's college for my undergraduate degree. One thing to note is that hostel accommodation for undergraduate women was available in women's colleges with a few seats in a postgraduate women's hostel. The co-ed colleges that had hostels had men's only hostels. Thus far more undergraduate male students stayed in on-campus accommodations than women. In the graduate master's program in India there was a higher relative percentage of women in theoretical physics streams like particle physics than in applied fields like electronics. I graduated from high school and college in india in the 80s. These incidents highlight attitudes from when I went to highschool and college in India. The job market has since grown and multinational companies and growing Indian companies in the technology sector have created many more STEM related jobs. Since then the biases may have become more subtle, but the attitude in the home is similar in that it is still assumed that STEM fields are the important ones to study. 

In the US I attended Washington University in St Louis for graduate studies. The enrolment of women in the physics graduate program was low with no women faculty outside the earth and planetary-science department. Now (from going through the listings on their website and the ratio of male:female graduate students is about 3:1). There are 33 current faculty listed out of whom there is 1 female Research Assistant Professor from Earth and Planetary Science, 1 female Research Associate Professor from earth and Planetary Science, 1 female Research Professor from Earth and Planetary Science, and 1 female Senior Lecturer. The rest are males.

While finishing my research at Washington University, I taught as a full time lab instructor (Sabbatical replacement position) at Saint Louis University for 1 semester (I had 5 male and 4 female teaching assistants mostly undergraduates). My Electricity and Magnetism, Mechanics, and Physics for pilots labs, had a very high male to female ratio. After graduate school I obtained a tenure-track faculty position at an HBCU college where I was an adjunct instructor when I was writing my thesis. I have gone through the stages of assistant, associate, and I am now a full professor. Our highest enrolment is of minority women.  In HBCUs the ratio of black women: black men is close to parity \citep{simms2014educational}.  

Many of our students come from underserved communities, and likely went to public schools with a high number of minorities. This community is underserved from elementary school onwards and so the issues are different from being a woman in STEM.  African-Americans have been kept from higher education longer than women in the US. Women were first admitted into Washington University in Saint Louis's medical school in 1918, while it was 1947 before the first African-Americans were allowed in the medical school. Washington University was fully desegregated only in 1954. There is a need for more science and math teachers in public schools and for after school programs that focus on math and science, and work in collaboration with colleges. This would be a good intervention program and should include competitions to encourage the students who have STEM aptitude that are similar to the athletics programs.

%https://www.statista.com/chart/4467/female-employees-at-tech-companies/

{\bf Christine Moran describes high school in Columbus, Ohio, USA, college at the MIT, graduate school at the University of Z\"{u}rich (UZH), postdoc work at Caltech, and work in industry.}
In my experience in high school, I wanted to fit in, so I often played up my absent minded professor side to the point of being thought a ``ditz" (a term primarily applied to women). Math and science always came easy to me, but I would hide the fact I often had the highest grade in the class. I had a small group of personal friends, most of whom were highly intelligent artists and activists, and none of whom were in my classes. I was in the robotics club, and had a part time job when I was 16 working at an engineering firm, but I thought math and science were boring because they were so easy and I wanted study philosophy then go to law school like my dad. 

I knew I could go to college for free if I went to an elite institution (Harvard, Stanford, MIT etc.) because of my single mother's finances and the financial aid these institutions provide. I read up on the typical profile of the student who was admitted to these colleges, and tried to fit it. Luckily I did well on standardized tests and although I found school boring, with a goal in mind it was easy to focus. I went to the information sessions for the elite colleges, and found most of them off putting in their elitism (I am not sure what I expected), except for MIT which I found down to earth, wacky and fun. The MIT spirit reminded me a lot of my friends and I thought it would be "funny" for a person who wanted to study philosophy and was known as a ditz to go to MIT. I applied early action, and was accepted. I often would get people reacting in surprise at the fact I would be attending MIT. One former physics teacher told me ``I hear they are letting in artsy people now". As a teenager I enjoyed surprising people or catching people off guard with my intelligence; as I became an adult this ``comedy" routine wore me down, because the humor came from people underestimating me, which was usually because of my gender, appearance, and affect. 

At MIT I found a place I felt home. Everyone took the same common mathematics and science core, which quickly lead me to realize these subjects could be challenging and rewarding and that I wanted to study more.  My admitted class was approximately 50\% women, and although my two majors (Physics, Computer Science and Engineering) were a smaller percentage, it was never small enough that I felt out of place. I think had I gone to any other college, I would not have entered the sciences and I'm thankful that my high school sense of humor brought me to MIT. At MIT I had many summer internships and chances to work in teams. In one summer internship, I traveled to a software engineering company in Norway where I was the first woman every hired in any department by the office in which I worked. This didn't feel off, although I recognize in retrospect I did a lot to seem ``one of the guys". After graduation, I worked at a local company called BBN Technologies doing machine translation research. I made an excellent salary and saved a lot of money. My team had several senior women on it, and was diverse in terms of age and cultural background. I ultimately decided to go to graduate school, and moved on to pursue my Master's and later Ph.D. at the University of Z\"urich (UZH).

It is not usual to receive funding as a Master's student in Europe where I had set my mind on going, but I was offered funding for 1 year (2/3 of my time) as a Master's student at the onset with the understanding the final 6 months of funding would follow at UZH. This clinched my decision to begin at the UZH. At UZH there was just one other woman studying for a Master's in my field at the same time. She and I bonded and did much of our studying together. I picked my Master's thesis with the understanding I would be working with the same supervisor as a Ph.D. During a meeting with this supervisor, it came out that he wanted me to work as the assistant to his secretary (who he happened to be dating) and do menial work for her in exchange for the final 1/3 of my funding, which I found demeaning given my skillset and refused. He offered the same ``opportunity" to the only other woman in my program, who despite my advice, took it. I am convinced this offer was related to my gender. At the same time, despite the fact I was doing well on my Master's thesis (I would receive the highest possible grade for its execution), he made clear that he would not be hiring me as a Ph.D. student. The same professor had also previously fired his only female Ph.D. student ever, and had hired another female Ph.D. physics student to clean his apartment (the rumor was, naked). The woman who took the job as the assistant to the secretary, struggled to be taken seriously in a scientific role because of her administrative duties, although working with her I can testify that she was equally talented as the men and myself in the department.

This sent me in a big scramble for money and into a crisis of grief and wondering what I would do next, I found it difficult to go to work. Another Professor in the department offered me a Ph.D. project, but as he was not the Professor I had set on working with and he had a rumor of being difficult with students and now I felt the environment to be toxic, I was unsure whether I wanted to stay. I began to eat into my USD savings to cover my living expenses, at the time that the USD was very weak, and quickly began consulting to make more money, while I finished my Master's. I went to a nearby University, the ETH, to work for a professor there for a few months while I decided whether I wanted to return. Ultimately I did return and accepted the Ph.D. offer of the second professor. I kept up my consulting projects, and took several unpaid leaves of absence to continue them, as well as to do scientific outreach projects. I graduated with my Ph.D. Many years later, my consulting work produced more than half a million dollars of revenue for me over my Ph.D., as I collected equity in the companies for which I worked, which when the companies were successful made me successful. I was happy I didn't take the job serving as a secretary's assistant. The woman who did was refused a Ph.D. position and ultimately dropped out of science.

Myself and my advisors assumed with my consulting work, that I would likely enter industry after graduation and I did not actively seek or receive counsel on postdoctoral fellowships. I did indeed begin to apply to industry, but as I had a budding interest in numerical relativity and had made some stabs at research in that direction and contacted professors abroad, decided to reach out to some of my contacts. One of them offered me a 1 year postdoctoral position at Caltech. I was delighted, and ended up deciding to move to LA to first work at SpaceX and later take the postdoctoral position. I was also excited because this supervisor was an avowed feminist and seemed to have many talented women working in his group. I moved to LA and worked for SpaceX for 3 months in an internship. There were very few women at SpaceX and none on my small team, although there were many among the intern class their numbers were diluted throughout the company. During my time at SpaceX I received active mentoring on submitting proposals and on potential postdoctoral fellowships to apply for by my future supervisor at Caltech, and I aided in submitting proposals and submitted my own. I started at Caltech, and the very next day heard that I had won one of the most prestigious: the NSF Astronomy and Astrophysics postdoctoral fellowship, which would fund me for a full 3 years at Caltech at roughly double my initial salary, with a mandate to do outreach work with a percentage of my time. I was delighted, and set off to make a big contribution and learn as much as I could about numerical relativity.

When I started at Caltech, within the first few weeks I began to hear rumors of my new advisor being difficult to work with. These rumors were never concrete and I am ashamed to say I didn't take them seriously. My Ph.D. advisor had been rumored to be difficult to work with, which I attributed to his hands off style, but it in the end worked out very well for me with my consulting schedule. I had the mistaking impression my postdoc advisor might have a similar situation, where some people had issues due to working styles, but I might find working with him just fine. I began my work, finally getting the chance to work with numerical relativity in depth, and collaborated mainly with another postdoc in the lab. The rumors intensified, and I was taken aside by a man in the department to explain that my advisor was especially bad news, and that he had tried to bully this man out of the field. The women in my advisors group began to leave, and they would make references to him making unreasonable demands. I began avoiding him entirely, preferring to work with the postdoc. I needed mentorship to progress as numerical relativity was a new branch of science to me, and when the postdoc left for a job in industry, and I found myself actively avoiding my postdoc advisor as it became clearer the extent of his bullying and harassment, and that it was targeting women, I thought about leaving and made plans to interview and take a sabattical position at the South Pole. My postdoc advisor was put on leave while a harassment investigation was underway. Shortly after that I left for Antarctica, hoping that the situation would be clearer when I returned a year later.

In Antarctica I ran the South Pole Telescope for 10 and 1/2 months together with a Ph.D. student, who also happened to be a woman. It was an amazing experience being in charge of such an impressive machine in a hostile environment and I was proud and comfortable to work with my colleague to do so. Partway through the year, Caltech wrote me that they had agreed with my former advisor to move the person who signed off on my grant to another professor in the department. When I returned from Antarctica, the situation at Caltech was still complicated. My former advisor was on leave, but was slated to return. He was prohibited from working with students, but could work with postdocs. I hadn't spoken to him for more than a year, he or Caltech had apparently requested the relationship be formally severed, and by now the rumors around his conduct were substantiated findings: he had engaged in gender based harassment of several of his students. The professor who signed off on my grant did very similar research to what I did in my Ph.D. and I worked to find common ground with my NSF proposal so that I could be more in line with his research. I then searched for a collaborator for my numerical relativity work, who ended up being a former Caltech postdoc with my former postdoc advisor, now working in the bay area. Collaboration at a distance was slow, and the numerical relativity work ground to a halt. I enjoyed and made progress with my research with the new advisor, but it wasn't what I had come to Caltech to do. I began to make preparations to leave my grant at Caltech early. I didn't feel like I could in good faith finish the research I set out to do on the grant, given the mess. I was able to fully execute on my outreach project as part of the grant, as well as publish a paper with my new advisor, before leaving the grant almost a year early to work at NASA JPL.

Without a system to report harassment, and with people being required to be silent about ongoing investigations, I felt that it became clear to me much too late that the problems students had with the advisor had nothing to due with differences of personality, and everything to do with the advisor. I wish I had known about these problems before coming to Caltech, and that afterwards I had not been so quick to map the problems to differences in personality. In reality, I should have switched research topics and advisors right away. 

My decision to leave my postdoc grant early was also impacted by my desire to start a family with my husband, who happened to have been the man who warned me my advisor was a bully. At this point I was in my early thirties, and my schedule dictated in another 2 to 3 years I may want to apply to a similar opportunity to the South Pole where being pregnant was not an option. So the next 2-3 years were ideal for pregnancy. But if we started a family while I was on my research grant, I would be searching for a new academic or industry job shortly before or after giving birth, and my maternity leave situation would be nebulous. My research grant offered unpaid leaves of up to three months, or the flexibility to work from home (some women did this throughout their maternity leave) at a self-decided pace while getting paid during this period. However, given how far I was behind on my grant work already, I couldn't see taking additional time from research as being possible, nor could I see going back full time to research days after giving birth as being practical. So I applied for and found a job at JPL. JPL has the advantage in that for many family leave options in the US you have to be working at an employer for more than a year to receive maternity leave outside of medical disability (approximately 6 weeks), but since JPL was managed by Caltech I would count as working there for more than a year were I to get pregnant, so I would be eligible for 12 weeks of leave (a combination of a percentage of my pay and unpaid leave) on top of approximately 6 weeks of disability. I went to my new postdoc advisor, shared with him the opportunity, and we agreed I would take it. I negotiated to keep my office at Caltech and to continue to finish some of the work we had begun since I returned from the South Pole. 

I had no idea how long it might take to get pregnant, but we began to hope our family dreams might come true within the 2-3 year window. Shortly after I started at JPL what my husband and I suspected was confirmed, I was pregnant. I chose to wait until the fetus was considered less likely to miscarriage to share the good news with my new boss. By that point, my husband had also switched jobs to join JPL as well. At first the HR told us that my husband and I would need to split a portion of our partially paid leave with each other because we shared the same employer. This is indeed the letter of the law in California. Later, we were told that Caltech had changed its policy to go above and beyond the law to allow us each a full 6 weeks of partially paid baby bonding leave. My husband and I sketched out a plan where we would each take 80 days off over the course of the baby's first year. My husband wasn't entitled to his leave until a year after his start date, as he didn't work for Caltech in his job immediately previous, so the majority of his leave would have to be taken after the baby was about 6 months old. However we were both glad we had stable jobs with leave provisions, as well as a wonderful daycare that we were admitted to for 3 days a week when the baby was 6 weeks or older. JPL is very diverse as far as the physics academia or tech industry standards go with respect to age, race, gender, and more, and is known to be highly supportive of families. We have found this to be the case thusfar. My group at JPL has a female principle investigator and a male manager, although the team has been for most of my time 100\% male otherwise. However, I work with a variety of missions, and most of them have heavy female representation. One particular mission I work with is majority female. I recall a meeting with 8+ women where a single man walked in late. I didn't not know who he was, so I introduced myself, he said his name was Guy, and I said "oh the token Guy." It's an honor and often relaxing to work with majority or all female teams, and one I have enjoyed rarely in my career.

I work out of my office at Caltech on research projects every other week now. Science will always be a part of my life. I have 13 refereed publications over the past 11 years, with more than 5000 citations in multiple academic disciplines. I can also publish as part of my job at JPL. I have a book the in the works, as well as several papers underway. My story became much more complicated and nebulous due to the intersection with male bullies, harassers, or other issues, but I have always found a creative way to make that into some sort of positive outcome. I hope I have learned more along the way so that I can be a better ally to those experience harassment. 

{\bf 
Anuja De Silva discusses her experience on single sex education in highschool in Sri Lanka and in the US at Mount Holyoke, graduate school at Cornell University, and industry work at IBM. She is also the mother of two children born while she was working in the semiconductor industry.}
In Sri Lanka, I attended a girls only high school where out of 400 in the graduating class more than 50\% are pursuing a career in the STEM fields.  I was among the 23  Mount Holyoke alumnae who went on to receive U.S. doctorates in life and physical sciences from 2000 to 2004, 22 of them are minority women, the highest number along first tier liberal arts colleges in the United States. As a minority Mount Holyoke graduate from class of 2004 who earned my PhD from Cornell University in 2009, I am proud to be featured in this statistic and still continue to work in the STEM field almost 10 years later.

The chemistry and chemical biology department of Cornell boasts more women than the physics department in keeping with the trends across universities. In 2003 the incoming class was about 35\% women and the numbers stayed consistent through graduation. The number of women in the department played a significant role in improving the graduate school experience for women. Friendship and support are key criteria that help  minorities succeed. I was fortunate to be a part of a research group led by an advisor who has consistently supported women graduates (our group was always 25-30\% women; 5-6 personnel) and hence kept faith in me after I ended up with unsatisfactory performance on certain classes that I had no study group support for. It was a turning point for me that bolstered more ambition and focus for me where as it could have easily turned into my giving up on the path towards a doctorate. During my thesis work, I learned that perseverance is key as I struggled for the first few years to attain meaningful results. But the support of my group members along with opportunities to interact with industry enabled me to stay motivated and focused. My graduate research was funded through an industry consortium through the Semiconductor Research Corporation (SRC), which included opportunities to attend conferences and well as an internship at IBM. This internship led to the start of my career as a post-doctoral associate at the IBM Almaden Research Center.

Working in the research and development area in a major tech company, diversity and inclusion of women is seen as a business imperative. During my tenure at IBM I have seen improvements to maternal leave policies and work life integration policies that enable both men and women to share the workload of child rearing. Due my background in women's only environments, I am more secure about my potential and contributions in a male dominated industry. But unlike in a women's ecosystem, I am also keenly aware of how much more assertive I need to be and how actively I need to seek recognition and promotion. As I enter the mid-career phase (around 10 years post graduate school), I am faced with the lack of women in technical and leadership roles first hand.  Women constitute about 25-35\% of these roles in most leading tech companies in the United States \citep{techcompanies}. As a mid-career woman my greatest challenge I face today to is to beat that statistic and continue my prominence in my technical field.

I do notice an active effort in corporate recruiting to emphasize the pipeline of women entering corporate research roles. Currently I am one of the senior technical staff who is also on a project lead role within my group. IBM is similar to other tech companies where the female technical staff is around 25-30\%.  Several initiatives in our division have gained management support and continue to bridge the gender disparity. They include gender bias training for all employees, an yearly all-day event that focuses on women'€™s empowerment, regular round tables with senior women engineers, and special assignments where women can spend up to 20\% time learning a new skill/working on a project with a mentor outside of their regular role. I have been in my current group for the past four years and I have seen an increase in contribution from women engineers in papers and patents published. As a senior woman in my group I have continued to seek out and engage women engineers and we have a high 50\% or more contribution from women in our published work. The women engineers also represent a variety of minorities (about 50\%).

Since my pregnancies were during my tenure in industry, I was eligible for paid maternity leave. With both children, I opted to stay home for 10 weeks post cesarean delivery and ease back to work. I was fortunate for the work life integration policies at my company which enabled me to take time off for doctor'€™s appointments and sudden child care emergencies. While I enlisted help from my mother, I also invested in paid child care, which was about 15-20\% of my salary. My childcare arrangements during work and work travel for conferences have been incurred on my personal income. I have considered them a longterm investment in myself and my family. I believe in consistently trying to find the best fit for my child as well as for my career.  The childcare costs tend to be the highest between ages 0 and 5. When children turn 5 in the United States, they start kindergarten. At that point parents who work full time will still incur costs of additional child-care at the beginning and at the end of the work day since a full day in school is only about 6 hours.  I am part of a dual income family. This has made a big difference in my ability to have two children relatively early in my career. For single income families, child rearing is a bigger financial strain. My current job also provides health benefits, lactation facilities and counselling which are necessities for working mothers.
\section{Conclusions}
The gender gap in STEM is attributed to the toxic climate that exists in male dominated fields. The attitude of colleagues is less overt against women and minorities than it was in the past, but it is still problematic and the numbers are changing slowly enough that gender parity will not be reached in fields like physics and engineering without intervention programs. Intervention programs have been shown to work. MIT succeeded in reaching gender parity in its undergraduate population in mechanical engineering. This was achieved through deliberate structural changes that included (1) reaching out to find and invite talented women and then hiring them as faculty instead of waiting for them to find the courage to apply to MIT, (2) promoting intervention programs like Women in Technology that invite talented highschool women to campus for a summer to take part in laboratories and courses, and (3) modernizing their curriculum in both content and pedagogy. 

To remove some of the toxicity of the environment, intervention programs have to go hand in hand with enforcing basic human rights that allow for sick leave and maternity/paternity leave for employees in industry and academia.  Not providing such leaves is an infringement on human rights that should no longer be the norm. It has been shown that providing employees with maternity leave and adequate day-care can be life-saving for both the mother and the child. In the academia, graduate students and postdoctoral students are temporary employees. They are expected to have worked abroad to qualify for faculty positions. Temporary employees do not have the same rights as staff, and if they hold a visa, they are not eligible for maternity and sick leave available to citizens even in countries where such leaves are the norm. Changes have to occur to support not only faculty in top universities to have families. All grants have to include provisions for basic human rights for graduate students and postdoctoral scholars so that universities do not rely on work-arounds that allow many women to fall through the cracks.

More than half of the women who obtain a STEM degree switch to other fields mid-career. To lose less talent, it is necessary to no longer see a degree as an end, and (1) create a network that connects students to the next level, (2) allow for maternity/paternity leave and return to work, (3) build support networks for students and employees and (4) actively train employees to facilitate a switch between projects when needed. 

The exposure of children to science is larger in Asia and Eastern Europe where the primary and secondary education focus on this end and families pressure both men and women to succeed in these fields. The result is that almost 50\% of women obtain science degrees, but the number of women who attain engineering and technology degrees is under 30\%. The latter number appears to be lower due to hidden and overt biases coupled with the expectation that men should be the providers of the family. We advocate gender neutral toys and the exposure of all children to science from the very beginning. The final goal, however, should be to have as many people as possible choosing what they would like to pursue and to be productive and innovative in their chosen fields. 

%right now STEM technical and engineering knowledge is mostly with males. 
In the end intervention programs must include the education of males as well. One reason all girls programs are going to only be part and not the whole solution is that at the moment the STEM technical knowledge and skills are mostly acquired by men. Women must be able to partake of the knowledge gained so far through interactions with both genders. However, women are often made to feel they are not capable of gaining this knowledge and there is hostility and ridicule when they try to gain it. This isolation can be removed in all female settings, however, the full benefit of access to the knowledge community will not be there. Hence changes in attitudes of both men and women are needed so that women feel comfortable as equal learning and working partners in STEM.

Ultimately, an increase in diversity where teams are not dominated by a single gender and by one or two national identities has been shown to lead to increased productivity, and to decrease the likelihood of abuse. To achieve diversity in STEM, we advocate exposing all children to science from kindergarten onwards and providing mentoring that evolves with age. Particular support is needed in the period post-graduate school. While attaining the PhD may seem as the end goal, it is just the beginning of a professional STEM career.
%Maternity leave is a basic human right that should not be violated so blatantly because its violation causes long-term health problems for both mothers and children. 
\section*{Acknowledgements}
We thank Kacey Bray Acquilano for support and for taking the time to provide us with Cornell physics data. We are very grateful to Dr. Mihai Bondarescu, Rowena Hazell and Prof. Ira Wasserman for suggestions and advice. 

%\section*{Supplemental Data}
% \href{http://home.frontiersin.org/about/author-guidelines#SupplementaryMaterial}{Supplementary Material} should be uploaded separately on submission, if there are Supplementary Figures, please include the caption in the same file as the figure. LaTeX Supplementary Material templates can be found in the Frontiers LaTeX folder.

%\section*{Data Availability Statement}
%The datasets [GENERATED/ANALYZED] for this study can be found in the [NAME OF REPOSITORY] [LINK].
% Please see the availability of data guidelines for more information, at https://www.frontiersin.org/about/author-guidelines#AvailabilityofData

%%%%%%%%%%%%%%%%%%%%%%



\bibliographystyle{mnras}
\bibliography{WomenInScience} % if your bibtex file is called example.bib


% Don't change these lines
\bsp	% typesetting comment
\label{lastpage}
\end{document}